\documentclass[sigconf]{acmart}
\usepackage{amsmath}
\usepackage{amsfonts}
\usepackage{amssymb}

\usepackage{balance}
\usepackage{multirow}
\usepackage{amsthm}
\usepackage{enumitem}
\usepackage{color}
\usepackage{booktabs}
\usepackage{multirow}

\usepackage{multirow}

\usepackage{graphicx}
\usepackage{float}
\usepackage{subfigure}
\usepackage{epstopdf}

\usepackage{bbding}

\usepackage{hyperref}
\usepackage{verbatim}

\newcommand{\ie}{\emph{i.e., }}
\newcommand{\eg}{\emph{e.g., }}

\usepackage{hyperref}

\AtBeginDocument{%
  }

\copyrightyear{2024}
\acmYear{2024}
\setcopyright{acmlicensed}\acmConference[RecSys '24]{18th ACM Conference on Recommender Systems}{October 14--18, 2024}{Bari, Italy}
\acmBooktitle{18th ACM Conference on Recommender Systems (RecSys '24), October 14--18, 2024, Bari, Italy}
\acmDOI{10.1145/3640457.3688118}
\acmISBN{979-8-4007-0505-2/24/10}

\acmSubmissionID{120}

\begin{document}

\title{Distillation Matters: Empowering Sequential  Recommenders \\ to Match the Performance of Large Language Models}

\author{Yu Cui}
\orcid{0009-0001-6203-3022}
\affiliation{%
  \institution{Zhejiang University}
  \city{Hangzhou}
  \country{China}}
\email{12321228@zju.edu.cn}

\author{Feng Liu}
\orcid{0009-0004-9265-9431}
\affiliation{%
  \institution{OPPO Research Institute}
  \city{Shenzhen}
  \country{China}}
\email{liufeng4hit@gmail.com}

\author{Pengbo Wang}
\orcid{0009-0003-1509-0256}
\affiliation{%
  \institution{University of Electronic Science and Technology of China}
  \city{Chengdu}
  \country{China}}
\email{2021080902021@std.uestc.edu.cn}

\author{Bohao Wang}
\orcid{0009-0006-8264-3182}
\affiliation{%
  \institution{Zhejiang University}
  \city{Hangzhou}
  \country{China}}
\email{bohao.wang@zju.edu.cn}

\author{Heng Tang}
\orcid{0009-0000-7933-8706}
\affiliation{%
  \institution{Zhejiang University}
  \city{Hangzhou}
  \country{China}}
\email{tangheng23@zju.edu.cn}

\author{Yi Wan}
\orcid{0009-0006-7876-7182}
\affiliation{%
  \institution{OPPO Research Institute}
  \city{Shenzhen}
  \country{China}}
\email{wanyi@oppo.com}

\author{Jun Wang}
\orcid{0000-0002-0481-5341}
\affiliation{%
  \institution{OPPO Research Institute}
  \city{Shenzhen}
  \country{China}}
\email{junwang.lu@gmail.com}

\author{Jiawei Chen}
\orcid{0000-0002-4752-2629}
\affiliation{%
  \institution{Zhejiang University}
  \department{The State Key Laboratory of Blockchain and Data Security}
  \city{Hangzhou}
  \country{China}}
\email{sleepyhunt@zju.edu.cn}
\authornote{Corresponding author.}

\renewcommand{\shortauthors}{Cui et al.}

\begin{abstract}

Owing to their powerful semantic reasoning capabilities, Large Language Models (LLMs) have been effectively utilized as recommenders, achieving impressive performance. However, the high inference latency of LLMs significantly restricts their practical deployment. To address this issue, this work investigates knowledge distillation from cumbersome LLM-based recommendation models to lightweight conventional sequential models. It encounters three challenges: 1) the teacher's knowledge may not always be reliable; 2) the capacity gap between the teacher and student makes it difficult for the student to assimilate the teacher's knowledge; 3) divergence in semantic space poses a challenge to distill the knowledge from embeddings.

To tackle these challenges, this work proposes a novel distillation strategy, DLLM2Rec, specifically tailored for knowledge distillation from LLM-based recommendation models to conventional sequential models. DLLM2Rec comprises: 1) \emph{Importance-aware ranking distillation}, which filters reliable and student-friendly knowledge by weighting instances according to teacher confidence and student-teacher consistency; 2)  \emph{Collaborative embedding distillation} integrates knowledge from teacher embeddings with collaborative signals mined from the data. Extensive experiments demonstrate the effectiveness of the proposed DLLM2Rec, boosting three typical sequential models with an average improvement of 47.97\%, even enabling them to surpass LLM-based recommenders in some cases.

\end{abstract}

\ccsdesc[500]{Information systems~Recommender systems}

\keywords{Sequential Recommendation, Large language Model, Knowledge Distillation}

\maketitle

\section{INTRODUCTION}
Large Language Models (LLMs) have showcased remarkable capabilities in content comprehension, generation, and semantic reasoning \cite{gpt3,achiam2023gpt,llama2}. Recently, LLMs have sparked a surge of interest within the domain of Recommender Systems (RS). Various research efforts have been devoted to harnessing LLMs to augment traditional recommendation models, serving as encoders for user/item features or as supplementary knowledge bases \cite{unisrec,hou2023learning,ren2023representation,tiger}. To fully exploit the reasoning ability of LLMs in RS, another line of research is to directly prompt or fine-tune LLMs to function as specialized recommenders. Owing to their inherent semantic reasoning capabilities, these LLM-based recommendation methods have achieved impressive performance. For instance, as shown in Table ~\ref{tab:inference time}, the representative model BIGRec \cite{bigrec} has demonstrated  approximately 41.44\% improvements on average over the state-of-the-art conventional sequential model (\ie DROS \cite{dros}) on the typical Amazon Games and Toys datasets.

Despite their effectiveness, LLM-based recommenders face serious inference inefficiency issues, posing substantial challenges to their practical applications. For example, as Table ~\ref{tab:inference time} shows, the widely used LLaMA2-7B model requires an astonishing three hours to perform a single inference for tens of thousands of users with 4x A800 GPUs. This inefficiency is exacerbated when scaling up to serve millions of users concurrently in practical industrial RS, where responses are required within seconds. This motivates a crucial research question: \textit{how can we maintain low inference latency as conventional recommenders while leveraging the excellent performance of LLM-based recommenders?}

To tackle this challenge, we propose employing knowledge distillation (KD) for acceleration --- \ie distilling the knowledge from a complex LLM-based recommendation model (teacher) to a lightweight conventional recommendation model (student). KD has been successfully applied in many domains \cite{hinton2015distilling,fitnets,unkd}, and has been proven effective in transferring knowledge from a large model to a smaller one. This strategy could capitalize on the effectiveness of LLM-based recommenders while maintaining low inference costs.   It also potentially integrates the capabilities of conventional models in capturing collaborative signals with the semantic reasoning prowess of LLMs. While the ideal is promising, distillation is non-trivial due to the fundamentally different mechanisms between the teacher and student models. LLMs primarily rely on content understanding and capturing semantic correlations for making recommendations, whereas conventional models depend on collaborative signals derived from users' historical behaviors. This divergence introduces several challenges: 

1) \textbf{Teacher Knowledge Reliability:} LLM-based models may not consistently outperform conventional models in all cases. Our empirical studies suggest that in over 30\% of cases, a conventional model could even outperform an LLM-based model, indicating that the knowledge from the teacher is not always reliable.  Moreover, the LLM may encounter the issue of notorious \textit{hallucination} and generate poor recommendations.

2) \textbf{Model Capacity Gap:} As demonstrated by recent work on KD \cite{cho2019efficacy}, the substantial difference in model size often makes it difficult for the student to assimilate the teacher's knowledge. Given the simple architecture of the conventional models, they may struggle to fully inherit the semantic reasoning ability of the teacher, and overloading the student with teacher knowledge might impair its own ability in collaborative filtering.

3) \textbf{Semantic Space Divergence:} Aligning embeddings for distillation, which has been demonstrated effective, presents significant challenges for this problem. LLMs model users/items primarily based on content, while conventional models derive users/items embeddings from collaborative signals. The vast semantic differences between these approaches mean that directly aligning their embeddings can be counterproductive, potentially disrupting the student's original embedding space and weakening its ability to capture collaborative signals.

To tackle these challenges, we propose DLLM2Rec, designed to effectively distill knowledge from LLM-based recommenders to conventional recommenders. DLLM2Rec contains: 

1) \textbf{Importance-aware ranking distillation.} Rather than directly aligning the ranking lists between the teacher and student, we highlight reliable and student-friendly instances for distillation by introducing importance weights. This approach evaluates the semantic similarity between the responses given by LLMs and the target positive items, with less similarity indicating lower response quality and suggesting such instances should be downweighted in distillation. Additionally, inspired by the ``wisdom of the crowd'', we leverage the model consistency between student and teacher to evaluate the importance of an instance, prioritizing instances where diverse models agree on higher item rankings. Such instances are also relatively easy and friendly to the student models, helping the student to assimilate the knowledge from the teacher.

2) \textbf{Collaborative embedding distillation.} To bridge the semantic gap between the embedding spaces of the teacher and student, we employ a learnable projector (\eg MLPs) to map original embeddings from teachers to the student's embedding space. Moreover, diverging from directly aligning the student embeddings with the teacher's projected embeddings, we introduce a flexible offset term that captures collaborative signals, further integrated with the teacher's projected embeddings to generate enriched student embeddings. This design effectively leverages the knowledge from the teacher while preserving its capacity to capture collaborative signals.

The main contributions of our work are summarized as follows:
\begin{itemize}[leftmargin=*]
\vspace{-0cm}
\item Highlighting the inference inefficiency issue of LLM-based recommendation model and advocating the use of knowledge distillation for acceleration.
\item Proposing DLLM2Rec which leverages importance-aware ranking distillation and collaborative embedding distillation to transfer reliable and student-friendly knowledge from LLM-based models to conventional recommendation models.
\item Conducting extensive experiments to demonstrate the effectiveness of DLLM2Rec, enabling lightweight conventional models to keep pace with sophisticated LLM-based models.
\end{itemize}

\begin{table}
  \caption{Recommendation performance and inference time-cost of BIGRec compared with DROS on Amazon Games and Toys datasets. Note that BIGRec is a typical LLM-based recommender with LlaMA-7B and DROS is a state-of-the-art sequential recommendation method. }
  \label{tab:inference time}
  \begin{tabular}{ccccc}
    \toprule
    Dataset&Model&HR@20&NDCG@20&Inference time\\
    \hline
    &DROS&0.0473&0.0267&1.8s\\
    &BIGRec&0.0532&0.0341&2.3$\times10^4s$\\
    \multirow{-3}{*}{Games}&\emph{Gain}&+12.47\%&+27.72\%&$-1.3\times10^6 \% $\\ \hline
    &DROS&0.0231&0.0144&1.6s\\
    &BIGRec&0.0420&0.0207&1.1$\times10^4$s\\
    \multirow{-3}{*}{Toys}&\emph{Gain}&+81.82\%&+43.75\%&$-6.8\times10^5 \% $\\
  \bottomrule
\end{tabular}
\vspace{-0.3cm}
\end{table}

\section{PRELIMINARIES}
In this section, we elaborate on sequential recommendation, and introduce BIGRec, a representative LLM-based recommender.

\subsection{Sequential Recommendation}
This work focuses on sequential recommendation, which has secured a pivotal role in various modern recommendation systems and attracted significant research interest. In a sequential recommender system with a user set $\mathcal{U}$ and an item set $\mathcal{I}$,  a user's historical interactions can be organized in chronological order $\mathbf s^t_u = (i_1, i_2, \ldots, i_{t-1})$ where $i_{k} \in \mathcal{I}$ represents the $k$-th item that the user $u$ interacted with. We remark that in this paper we may
shorten the notation $\mathbf s^t_u$ as $\mathbf s$ for clear presentation.
The task of sequential recommendation is to predict the next item $i_{t}$ that the user is likely to interact with. 

\textbf{Sequential Recommendation Model.} Existing models in this domain primarily adopt representation learning paradigms. These methods first utilize an item encoder to map items' features $x_i$ (\eg IDs, titles) into their representations $\mathbf e_i$: 
\begin{equation}
\setlength{\abovedisplayskip}{3pt}
\setlength{\belowdisplayskip}{3pt}
  \begin{aligned}
\mathbf e_i=\text{ItemEncoder}(x_i)
  \end{aligned}
\end{equation}
where ItemEncoder(.) can be implemented by various architectures, \eg an embedding layer to encode item ID or BERT \cite{devlin2018bert} to encode item text. 

With the item embeddings, user behaviors can be further encoded by a sequential encoder:
\begin{equation}
\setlength{\abovedisplayskip}{3pt}
\setlength{\belowdisplayskip}{3pt}
  \begin{aligned}
\mathbf e_{\mathbf s}=\text{SeqEncoder}(\mathcal Z_{\mathbf s})
  \end{aligned}
\end{equation}
where $\mathcal Z_{\mathbf s}$ denotes the encoded item embedding sequence of $\mathbf s$, \ie  $\mathcal Z_{\mathbf s}=(\mathbf e_{i_1}, \mathbf e_{i_2}, \ldots, \mathbf e_{i_{t-1}})$. SeqEncoder(.) represents the sequence encoder and can be implemented by GRUs \cite{hidasi2015session}, Transformer \cite{kang2018self}, or other architectures. 

Given the sequence and item embeddings, the final prediction $\hat y_{\mathbf si}$ can be generated via the dot product \cite{ogita2005accurate} or MLPs \cite{pinkus1999approximation}, which is then utilized to retrieve recommendations. Let $i^*_{\mathbf s}$ (shorten as $i^*$) be the ground-truth item of the sequence $\mathbf s$ that the user $u$ will interact with at the next step, the model can be trained via various losses, \eg binary cross-entropy \cite{ruby2020binary}:
\begin{equation}
\setlength{\abovedisplayskip}{3pt}
\setlength{\belowdisplayskip}{3pt}
  \begin{aligned}
      \mathcal{L}_r=-\sum_{\mathbf s\in \Gamma }\bigg (\log\sigma(\hat y_{\mathbf si^*})+\sum_{{j\in O^{-}}}\log(1-\sigma(\hat y_{\mathbf sj}))\bigg )
  \end{aligned}
  \end{equation}
where $\sigma(.)$ denotes the Sigmoid function; $\Gamma$ denotes the set of sequences used for model training; and $O^{-}$ denotes the set of sampled negative items.

\subsection{Brief on BIGRec}
Recently LLM-based recommendation attracts great attention. Predominantly, this body of work formulates the recommendation task using natural language prompts and employs large language models to generate personalized recommendations \cite{hou2024large}. This study simply take the representative model BIGRec \cite{bigrec} for empirical analysis. The selection of BIGRec is justified not only by its availability as an open-source tool but also by its demonstrated effectiveness. Furthermore, BIGRec embodies the fundamental elements of LLM-based recommendation and many methods can be considered as further extensions of such paradigm \cite{sun2024large,wang2024can}. It is also noteworthy that BIGRec has been employed by recent studies \cite{lin2024data,grounding2} as a representative model for analysis.

\begin{figure*}[htbp]
    \centering 
     \setlength{\abovecaptionskip}{-0ex}
    \includegraphics[width=1\textwidth]{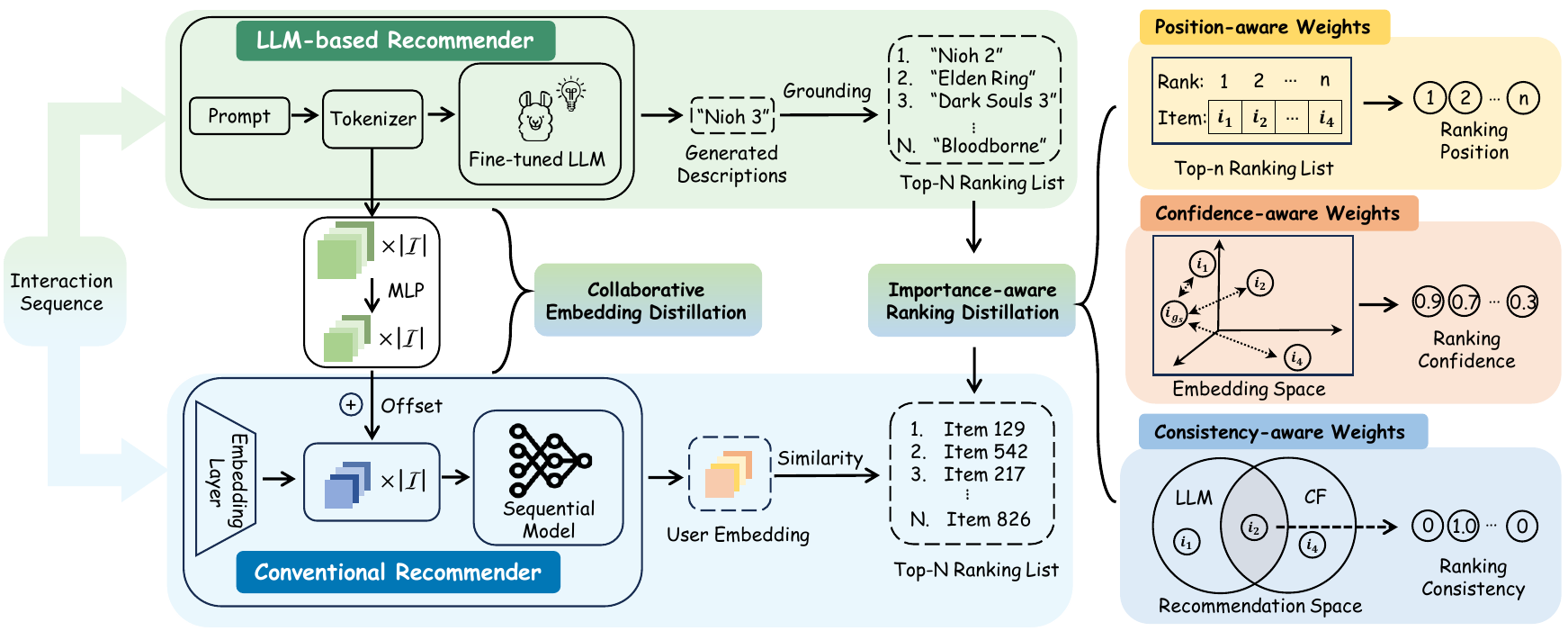}
    \caption{Illustration the proposed DLLM2Rec that distills the knowledge from the LLM-based recommenders to the conventional recommenders, with leveraging importance-aware ranking distillation and collaborative embedding distillation.} 
    \label{fig:Framework} 
\end{figure*}

To be specific, BIGRec organizes users' historical behaviors in natural language and employs instruction-tuning to fine-tune LLMs, as illustrated in Figure \ref{fig:Framework}. During the inference stage, BIGRec generates item descriptions (\eg titles) for recommendations. Considering that these descriptions may not always correspond to existing items, BIGRec incorporates a grounding paradigm that matches generated item descriptions to existing items based on content similarity. Formally, let $\mathbf{z}_{g_{\mathbf s}}$ and $\mathbf{z}_i$ represent the token embeddings of generated descriptions and the descriptions of item $i$, respectively. BIGRec computes their L2 distance for grounding as follows:
\begin{equation}
\setlength{\abovedisplayskip}{3pt}
\setlength{\belowdisplayskip}{3pt}
  \begin{aligned}
d_{\mathbf si}=||\mathbf z_{g_{\mathbf s}}-\mathbf z_i||^2
  \end{aligned}
\end{equation}
Based on $d_{\mathbf si}$, BIGRec ranks items and retrieve the K-nearest items as recommendations.

\section{METHODOLOGY}
In this section, we first outline the challenges associated with distilling knowledge from Large LLM-based recommendation models to conventional models (subsection \ref{section:Motivations}). Following this, we delve into the specifics of our proposed DLLM2Rec (subsections \ref{section:Proposed Distillation Strategy: DLLM2Rec}).

\subsection{Motivations}
\label{section:Motivations}
In this subsection, we aim to conduct a thorough analysis to elucidate the challenges of distillation, thereby motivating our proposed method. These challenges can be categorized into three aspects:

\textbf{Teacher Knowledge Reliability.} By examining the recommendation results from a typical LLM-based model, BIGRec, and a state-of-the-art sequential model, DROS, on three real-world datasets, we empirically discover that the LLM-based model may not consistently surpass the conventional model in all test cases. In fact, as depicted in Table \ref{tab:great}, BIGRec may underperform DROS in over 30\% of cases across all three datasets. This observation implies that the teacher's knowledge may not always be reliable and could potentially be detrimental. The reliability of the teacher's knowledge in distillation must be validated.
 \begin{table}[tbp]
 \vspace{-0em}
  \caption{The ratio of cases where BIGRec outperforms DROS to cases where BIGRec underperforms DROS on NDCG@20. }
  \vspace{-0.3cm}
  \label{tab:great}
  \begin{tabular}{ccc}
    \toprule
    Dataset& Condition	&  Relative Ratio\\
    \hline
                            &BIGRec > DROS & 53.90\%\\               
    \multirow{-2}{*}{Games} &BIGRec < DROS & 46.10\%\\  \hline
                                &BIGRec > DROS & 40.90\%\\
    \multirow{-2}{*}{MovieLens} &BIGRec < DROS & 59.10\%\\  \hline
                           &BIGRec > DROS & 66.67\%\\
    \multirow{-2}{*}{Toys} &BIGRec < DROS & 33.33\%\\
    \bottomrule
\end{tabular}
\vspace{-0cm}
\end{table}

\begin{table}[t]
  \caption{Ratio of overlapped items in Top-20 recommendations between BIGRec and DROS. Additionally, we present the percentage of these items that are actual hits. For comparative analysis, we detail the values specific to items unique to either BIGRec's or DROS's recommendations. 
  }
  \vspace{-0.3cm}
  \label{tab:consis}
  \begin{tabular}{cccc}
    \toprule
    Dataset&Rec. Space &Items Ratio	& Hit Items\\
    \hline
    &BIGRec only&96.01\%&0.21\%\\
    &DROS only&96.01\%&0.18\%\\
    \multirow{-3}{*}{Games}&Overlapped&3.99\%&1.61\% \\ \hline
    &BIGRec only&95.94\%&0.19\%\\
    &DROS only&95.94\%&0.35\%\\
    \multirow{-3}{*}{MovieLens}&Overlapped&4.06\%&2.16\% \\ \hline    
    &BIGRec only&98.95\% &0.17\%\\
    &DROS only&98.95\% &0.08\% \\
    \multirow{-3}{*}{Toys}&Overlapped&1.05\% &3.74\% \\
  \bottomrule
\end{tabular}
\end{table}

\textbf{Model Capacity Gap.}  Recent research suggests that the performance of a student model diminishes as the gap in size between the teacher and student models increases \cite{huang2022knowledge}. This challenge is even more pronounced in our scenario, where the student model comprises a million-level parameters while the teacher model has billion-level parameters. Additionally, the teacher and student models employ fundamentally different recommendation mechanisms. We notice a significant discrepancy in their recommended items --- the average number of overlapped items in their Top-20 recommendations is less than 3.15\% across the three datasets as shown in Table \ref{tab:consis}. It is implausible to expect the student to fully assimilate the teacher's knowledge and fully inherit the teacher's ability on semantic reasoning. Overloading the student with the teacher's knowledge may even impair the student's inherent capacity to capture collaborative signals. Our empirical study, as shown in Table \ref{tab:performance}, demonstrates that existing knowledge distillation strategies usually yield limited improvements and can sometimes even be counterproductive. Thus, the development of a distillation strategy that is friendly to the student model is crucial.

\textbf{Semantic Space Divergence.} It is noteworthy that LLM-based models characterize users/items mainly based on their contents, while conventional models derive users/items embeddings mainly from collaborative signals. It means the teacher and student adopt entirely different semantic frameworks. Blind alignment of their semantic spaces for distillation could prove counterproductive. As observed in Table \ref{tab:performance}, two representative knowledge distillation methods, Hint \cite{fitnets} and HTD \cite{htd-21}, which distill through embeddings, often perform worse than the original student model without knowledge distillation. While embedding distillation has proven effective in many domains, it should be specifically designed for this task.

\subsection{Proposed Distillation Strategy: DLLM2Rec}
\label{section:Proposed Distillation Strategy: DLLM2Rec}
In order to address the aforementioned challenges, this work proposes DLLM2Rec, with leveraging importance-aware ranking distillation and collaborative embedding distillation.

\subsubsection{Importance-aware Ranking Distillation.} This module builds upon the conventional ranking distillation \cite{rd} while additionally introducing importance weights to emphasize reliable and student-friendly instances. Specifically, we employ the following distillation loss:
\begin{equation}
\setlength{\abovedisplayskip}{3pt}
\setlength{\belowdisplayskip}{3pt}
  \begin{aligned}
      \mathcal{L}_d=-\sum_{\mathbf s \in \Gamma }\sum_{i \in \mathcal O^T }w_{\mathbf si}\log\sigma(\hat y_{\mathbf si})
  \end{aligned}
  \end{equation}
where $\mathcal O^T$ denotes the Top-K recommendations returned by the teacher model, and $w_{si}$ denotes the distillation weight. In this work, we choose $K=10$ as a default value, but this can be tuned for optimal performance. The objective is straightforward  --- we select the highly ranked items from the teacher as a positive to guide the learning of the student, so that these candidate items can also be recommended by the student. However, given that the teacher's recommendations may not always be beneficial, we introduce an importance weight that considers the following three aspects:

1) \textbf{Position-aware weights $w^p_{\mathbf si}$.}  Inheriting from \cite{rd}, ranking positions are also considered in DLLM2Rec. The motivation is from the ranking alignment that we would like to push a candidate item higher if the item also occupies a higher position in the teacher's ranking list. Formally, we use:
\begin{equation}
\setlength{\abovedisplayskip}{3pt}
\setlength{\belowdisplayskip}{3pt}
  \begin{aligned}
  w^p_{\mathbf si} \propto \exp(-r_i/\beta)
\end{aligned}
\end{equation}
where $r_i$ denotes the position of item $i$ in the ranking list returned by the teacher, and $\beta$ is the hyperparameter adjusting the shape of the weight distribution. 

2) \textbf{Confidence-aware weights $w^c_{\mathbf si}$.}  Given the importance of extracting reliable teacher knowledge, we leverage $w^c_{\mathbf si}$ to indicate reliability. Specifically, we measure the quality of descriptions generated by LLMs by assessing the content distance between the generated descriptions and the content of the ground-truth item:
\begin{equation}
\setlength{\abovedisplayskip}{3pt}
\setlength{\belowdisplayskip}{3pt}
  \begin{aligned}
  & w^c_{\mathbf si} \propto \exp(-d_{\mathbf si^*}/\beta) \\
    & d_{\mathbf si^*}=||\mathbf z_{{g_{\mathbf s}}} -\mathbf z_{i^*}||^2
\end{aligned}
\end{equation}
where $d_{\mathbf si^*}$ measures the embedding distance between the generated item description $\mathbf z_{g_{\mathbf s}}$ and the target ground-truth item $\mathbf z_{i^*}$, where the embeddings can be generated via LLMs encoder. A smaller distance suggests a higher quality of the generated description as it aligns more closely with the targets. Conversely, a larger gap suggests lower confidence, indicating that LLMs may risk generating incorrect or nonsensical information.

3) \textbf{Consistency-aware weights $w^s_{\mathbf si}$.} Inspired by the ``wisdom of the crowd'', we use model consistency between the student and teacher to assess the importance of an instance.  As suggested by recent work on bagging \cite{du2020agree,you2017learning}, when diverse models reach a consensus on one prediction, its reliability increases. In RS, our empirical studies (Table \ref{tab:consis}) also shows that the items that are concurrently recommended by teacher and student are more likely to be positive. This insight allows us to formulate consistency-aware weights as follows:

\begin{equation}
\setlength{\abovedisplayskip}{0cm}
\setlength{\belowdisplayskip}{0.3cm}
  w^o_{\mathbf si}=\left\{ {\begin{array}{*{20}{c}}
    {1,\quad i \in \mathcal O^T \cap  \mathcal O^S }\\
    {0,\quad i \notin \mathcal O^T \cap  \mathcal O^S }
    \end{array}} \right.
\end{equation}
where $\mathcal O^T$ and $\mathcal O^S$ denote the sets of Top-K recommendation items returned by the teacher and student, respectively. We assign higher weights to those overlapping items (\ie $i \in \mathcal O^T \cap  \mathcal O^S $).

Another advantage for up-weighting those overlapped items is that  they are relatively easy and friendly for student learning. By examining the gradient of the distillation loss:
\begin{equation}
\setlength{\abovedisplayskip}{3pt}
\setlength{\belowdisplayskip}{3pt}
  \begin{aligned}
    \frac{\partial \mathcal{L}_d }{\partial \hat y_{\mathbf si}} = w_{\mathbf si}\sigma(-\hat y_{\mathbf si})
\end{aligned}
\end{equation}
it is evident that instances with larger $\hat y_{\mathbf si}$, \ie  higher positions in the student ranking lists, will have smaller gradient magnitudes. This suggests that higher-ranked instances are more easily assimilated by the student model, as the student does not require to make extensive change. Upweighting these instances makes the knowledge distillation process more conducive to student learning.

We integrate these three aspects into the ranking distillation with a simple linear combination:
\begin{equation}
\setlength{\abovedisplayskip}{3pt}
\setlength{\belowdisplayskip}{3pt}
  w_{\mathbf si}=\gamma_p \cdot w^p_{\mathbf si} + \gamma_c \cdot w^c_{\mathbf si} + \gamma_o \cdot w^o_{\mathbf si}
  \label{equation:wd}
\end{equation}
where $\gamma_p, \gamma_c, \gamma_o$ are hyperparameters balancing their contributions. The ultimate objective of our DLLM2Rec is:
\begin{equation}
\setlength{\abovedisplayskip}{3pt}
\setlength{\belowdisplayskip}{3pt}
 \mathcal{L}=\mathcal{L}_r+\lambda_d\mathcal{L}_d
\end{equation}
where $\lambda_d$ balances the contributions from the recommendation and distillation losses.

Interestingly, some recent work \cite{rd,dcd} consider to up-weight the instances which has larger ranking discrepancy between student and teacher. This strategy is ineffective in this task, as it would increase the distillation unreliability and difficulty. Our DLLM2Rec adopts contrary strategy and would yield better performance as demonstrated in our experiments. 

\subsubsection{Collaborative Embedding Distillation.} Embedding distillation has proven effective in many domains, yet it requires careful design in this context, given that the teacher and student possess quite different semantic spaces.
To tackle this, we adopt a collaborative paradigm. Specifically, we first employ a learnable projector (\eg MLPs) to map original item embeddings from the teacher to the student's embedding space to bridge the semantic gap:
\begin{equation}
\setlength{\abovedisplayskip}{3pt}
\setlength{\belowdisplayskip}{3pt}
  \begin{aligned}
\mathbf z^p_{i}= g(\mathbf z_i)
\end{aligned}
\end{equation}
where $\mathbf z_i$ denotes the textual embedding of item $i$ encoded by the LLM-based recommender; $g(.)$ denotes a learnable projector function, which can be implemented via MLPs.

we further introduce a flexible offset term $\mathbf b_i$ for each item, which is integrated with the teacher's projected embeddings to generate enriched student embeddings:
\begin{equation}
\setlength{\abovedisplayskip}{3pt}
\setlength{\belowdisplayskip}{3pt}
  \begin{aligned}
\mathbf e^{new}_{i}= f(\mathbf z^p_{i},\mathbf b_{i})
\end{aligned}
\end{equation}
where $\mathbf b_{i}$ is a learnable vector designed to capture the collaborative signals from user behavior data. $f(.)$ denotes a function combining the distilled information from the teacher and the information mined from the data.  $f(.)$ can be implemented via various ways, \eg concatenate, MLPs. In our experiments, we find that simple linear combinations suffice to yield satisfactory performance. Such collaborative approach allows our model to leverage the powerful knowledge from the teacher while preserving its capacity to capture collaborative signals.

\textbf{Remarkably, the distilled student is as efficiency as the original one during the inference}. It only requires to leverage LLMs in model training, while directly utilize the well-trained and empowered student model for online service.

\section{EXPERIMENTS}
Our experiments address the following research questions:

\begin{itemize}[leftmargin=*]
  \item $\mathbf{RQ1:}$ Does DLLM2Rec outperform existing distillation strategies? How does the empowered student model perform?
  \item $\mathbf{RQ2:}$ What are the impacts of different components of DLLM2Rec on its performance?
  \item $\mathbf{RQ3:}$ How do hyperparameters influence DLLM2Rec?
  \end{itemize}

\subsection{Experiment Settings}

\subsubsection{Datasets}
Three conventional datasets: \textit{Amazon Video Games}, \textit{MovieLens-10M}, and \textit{Amazon Toys and Games} were utilized in our experiments \footnote{\url{https://jmcauley.ucsd.edu/data/amazon/}}\textsuperscript{,}\footnote{\url{https://grouplens.org/datasets/movielens/10m/}}. These datasets include user behavior sequences and item content. For fair comparisons, we closely adhered to the preprocessing methods used in recent work \cite{dros,bigrec}. We organized the interaction sequences in ascending order of timestamps to partition each dataset into training, validation, and testing sets with ratios of 8:1:1.  Given that MovieLens-10M contains an excessive number of sequences, which could not be processed by LLM-based recommenders, we sampled 100,000 sequences for training and 10,000 for testing, the sampling strategy also adopted by \cite{bigrec}.
The dataset statistics are presented in Table~\ref{tab:datasets}.


\begin{table}
  \caption{Statistics of the datasets.}
  \vspace{-0.3cm}
  \label{tab:datasets}
  \begin{tabular}{ccccc}
    \toprule
     \textbf{Datasets} & \textbf{Games} & \textbf{MovieLens} & \textbf{Toys}\\
    \hline    
    \#Users & 55,223 & 69,878 & 19,412\\
    \#Items & 17,408 & 10,681 & 11,924\\
    \#Interactions & 497,577 & 1,320,000 & 167,597\\
    Density & 0.05176\% & 0.1769\% & 0.07241\%\\
  \bottomrule
\end{tabular}
\vspace{-0.5cm}
\end{table}

\subsubsection{Baselines}

The following strategies are compared: 

\textbf{Knowledge Distillation Strategies for Recommendation: }  \textbf{Hint} \cite{fitnets}, \textbf{HTD} \cite{htd-21} are two representative methods that distill knowledge through teacher embeddings;  \textbf{RD}  \cite{rd}, \textbf{CD} \cite{cd}, \textbf{RRD} \cite{rrd}, \textbf{DCD} \cite{dcd} and \textbf{UnKD} \cite{unkd} distill information from the teachers' ranking lists. Readers may refer to the related work for more details about these strategies.

\textbf{LLM-enhanced Recommendation Methods:} We selected KAR \cite{xi2023towards} and LLM-CF \cite{sun2024large} for comparisons as they are open-sourced, closely related, and state-of-the-art.  \textbf{KAR} leverages LLMs as knowledge base to enhance the profile of users and items, while \textbf{LLM-CF} uses LLMs to generate a base of chains of thought, which are further retrieved to enhance sequential models.

For fair comparisons, we integrated these methods into three representative sequential models: GRU4Rec \cite{hidasi2015session}, SASRec \cite{kang2018self}, and DROS \cite{dros}, which are either well-known or state-of-the-art. For the teacher model, we consistently used the LLM-based model BIGRec \cite{bigrec}, due to its availability as an open-source tool and its demonstrated effectiveness.

\subsubsection{Evaluation Metrics}
We employed two widely-used metrics HR@K and NDCG@K to evaluate performance. Here we simply set K to 20 as recent work \cite{lightgcn}, and observed similar results with other choices of $K$.

\subsubsection{Implementation Details}

All methods are implemented with PyTorch and run on 4 Nvidia A800 GPUs. We set $\beta=1.0, \gamma_p=0.3, \gamma_c=0.5, \gamma_o=0.1$ across all datasets, as these settings were found sufficient to generate good performance, although fine-tuning could further enhance model performance. The influence of these hyperparameters on model performance is also presented in Figure \ref{fig:Sensitivity}. Adam \cite{adam} was used as the optimizer with a tuned learning rate of 0.001, a batch size of 256, and weight decay tuned in \{1e-4, 1e-5, 1e-6, 1e-7, 0\}, $\lambda_d$ in \{0.1, 0.2, ..., 1.0\}. We set the embedding size to 64 and the dropout ratio to 0.1. Our code is available on \footnote{\url{https://github.com/istarryn/DLLM2Rec}}.

For all compared methods, we closely followed the settings suggested by their original papers. We also finely tuned their hyperparameters to ensure their optimum. Specifically, for BIGRec, we implemented it with LLaMA2 \cite{llama2} as suggested by the authors.

\subsection{Performance Comparison (RQ1)}

The overall experimental results are presented in Table~\ref{tab:performance}.

\begin{table*}[htbp]
\vspace{-0.3cm}
  \caption{Performance comparisons of DLLM2Rec with existing KD methods and LLM-enhanced strategies. The best performance is bold while the runner-up is underlined. \emph{Gain.S} denotes the improvement of DLLM2Rec over the student; while \emph{Gain.B} denotes the improvement of DLLM2Rec over the best baseline. }
  \vspace{-0.3cm}
  \label{tab:performance}
  \begin{tabular}{cccccccc}
    \toprule
\multirow{2}{*}{Backbone} &\multirow{2}{*}{Model}       & \multicolumn{2}{c}{ Games} & \multicolumn{2}{c}{ MovieLens} & \multicolumn{2}{c}{ Toys}\\ 
\cline{3-8} 
                                 &                   & HR@20   & NDCG@20      & HR@20  & NDCG@20        & HR@20 & NDCG@20       \\
\hline
                        Teacher  & BIGRec          & 0.0532  & 0.0341       & 0.0541 & 0.0370         & 0.0420 & 0.0207      \\        
\hline
                                 & +None          & 0.0305  & 0.0150       & 0.0608 & 0.0236         & 0.0172  & 0.0081           \\
                                 & +Hint          & 0.0284  & 0.0120       & 0.0646 & 0.0240   & 0.0128 & 0.0058                 \\
                                 & +HTD             & 0.0299  & 0.0128       & 0.0578 & 0.0229   & 0.0155 & 0.0062                 \\
                                 & +RD              & 0.0398  & 0.0177       & 0.0667 & 0.0254   & 0.0157 & 0.0076                 \\
                                 & +CD              & 0.0306  & 0.0149       & \underline{ 0.0699} & 0.0256 & 0.0126 & 0.0052           \\
                                 & +RRD             & 0.0359  & 0.0163       & 0.0657 & 0.0243         & 0.0215 & 0.0097           \\
                                 & +DCD             & \underline{ 0.0427}  & \underline{0.0190}       & 0.0666 & \underline{ 0.0263}         & \underline{ 0.0262} & \underline{ 0.0114}           \\
                                 & +UnKD            & 0.0370  & 0.0170       & 0.0607 & 0.0226         & 0.0235 & 	0.0114           \\
                                 & KAR               & 0.0307  & 0.0149       & 0.0603 & 0.0229         & 0.0184 & 0.0079         \\
                                 & LLM-CF            & 0.0393	& 0.0174       & 0.0677 & 0.0246         & 0.0132 & 0.0058         \\
                                 & +DLLM2Rec          & \textbf{0.0446} & \textbf{0.0205}  & \textbf{0.0815} & \textbf{0.0308} & \textbf{0.0281} & \textbf{0.0118}          \\
                                 \cline{2-8}
                                 & \emph{Gain.S}  & +46.17\% & +36.94\%   & +34.05\% & +30.43\%    & +63.88\% & +42.18\%               \\
\multirow{-13}{*}{ GRU4Rec}      & \emph{Gain.B}  & +4.56\% & +7.64\%   & +16.60\% & +16.80\%    & +7.40\% & +1.27\%               \\ 
\hline
                                 & +None          & 0.0346  & 0.0190       & 0.0626  & 0.0228        & 0.0207 	& 0.0130            \\
                                 & +Hint          & 0.0358  & 0.0151       & 0.0576 & 0.0216   & 0.0242 & 0.0103                 \\
                                 & +HTD             & 0.0343  & 0.0152       & 0.0569 & 0.0214   & 0.0209 & 0.0097                 \\
                                 & +RD              & 0.0513  & 0.0225  & 0.0778  & \underline{ 0.0310}   & \underline{ 0.0397} & 0.0164           \\
                                 & +CD              & 0.0396  & 0.0231 & 0.0712  & 0.0265        & 0.0232 & 0.0151            \\
                                 & +RRD             & 0.0479  & 0.0202       & 0.0633  & 0.0244        & 0.0325 & 0.0158            \\
                                 & +DCD             & 0.0455  & 0.0211       & 0.0723  & 0.0275        & 0.0375 & \underline{ 0.0175}            \\
                                 & +UnKD            & 0.0447  & 0.0219       & 0.0667  & 0.0247        & 0.0335 & 0.0174            \\
                                 & KAR               & 0.0381	  & 0.0198       & 0.0565	 & 0.0221         & 0.0215	 & 0.0131         \\
                                 & LLM-CF            &\underline{  0.0559}     & \underline{ 0.0251}       & \underline{ 0.0837}	 & 0.0295         & 0.0335 & 0.0152         \\
                                 & +DLLM2Rec          & \textbf{0.0600} & \textbf{0.0262} & \textbf{0.0840} & \textbf{0.0323}  & \textbf{0.0409} & \textbf{0.0177}         \\
                                 \cline{2-8}
                                 & \emph{Gain.S}  & +73.55\% & +38.25\%   & +34.19\% & +41.91\%    & +97.68\% & +36.38\%               \\
\multirow{-13}{*}{ SASRec}       & \emph{Gain.B}  & +7.36\% & +4.40\%   & +0.36\% & +4.34\%    & +3.02\% & +1.19\%               \\ 
\hline
                                 & +None          & 0.0473  & 0.0267       & 0.0852  & 0.0363        & 0.0231 	& 0.0144            \\
                                 & +Hint          & 0.0531  & 0.0240       & 0.0791 & 0.0306   & 0.0302 & 0.0135                 \\
                                 & +HTD             & 0.0489  & 0.0238       & 0.0722 & 0.0289   & 0.0275 & 0.0137                 \\         
                                 & +RD              & 0.0585  & \underline{ 0.0310} & 0.0950 & \underline{ 0.0383}    & 0.0424 & \underline{ 0.0220}          \\
                                 & +CD              & 0.0474  & 0.0270       & 0.0802  & 0.0336        & 0.0238 & 0.0156            \\
                                 & +RRD             & 0.0590  & 0.0293       & 0.0788  & 0.0338        & 0.0424 & 0.0212            \\
                                 & +DCD             & 0.0531  & 0.0273       & 0.0821  & 0.0348        & \underline{ 0.0432} & 0.0211            \\
                                 & +UnKD            & 0.0448  & 0.0209       & 0.0728  & 0.0297        & 0.0375 & 0.0195            \\
                                 & KAR               & 0.0586  & 0.0318       & 0.0859 & 0.0352         & 0.0255 & 0.0156         \\
                                 & LLM-CF            & \underline{ 0.0635}  & 0.0293       & \underline{ 0.0963} & 0.0351         & 0.0385 & 0.0178         \\
                                 & +DLLM2Rec          & \textbf{0.0751} & \textbf{0.0331}  & \textbf{0.1063} & \textbf{0.0437}   & \textbf{0.0463} & \textbf{0.0225}        \\
                                 \cline{2-8}
                                 & \emph{Gain.S}  & +58.77\% & +23.90\%   & +24.77\% & +20.41\%    & +100.43\% & +56.35\%               \\
\multirow{-13}{*}{ DROS}         & \emph{Gain.B}  & +18.27\% & +4.03\%   & +10.38\% & +14.24\%    & +7.07\% & +2.16\%               \\ 
\hline
\end{tabular}
\vspace{-0.3cm}
\end{table*}

\textbf{Comparing with students.}  The improvements brought by DLLM2Rec are impressive, achieving an average improvement of 47.97\% over the original students across three datasets and two metrics. Furthermore, these improvements are consistent under all conditions. These results clearly validate the effectiveness of DLLM2Rec in distilling useful knowledge from the teacher to enhance the student models.

\textbf{Comparing with exising KDs.} DLLM2Rec consistently outperformed all KD baselines across all datasets and metrics. This result clearly validates the effectiveness of DLLM2Rec, with leveraging reliable and student-friendly distillation strategies. We also observed that some KD methods, \eg Hint and HTD, showed a negative impact on recommendation performance compared to the original student model. This could be attributed to the large semantic gap between the teacher and student models. Additionally, some advanced KD methods like UnKD, DCD, HTD, and RRD may be inferior to the basic RD in some scenarios. This could be attributed to these advanced KD methods adopting more complex distillation strategies, increasing the difficulty for the student to digest the knowledge. 

\textbf{Comparing with exising LLM-enhanced methods.}  DLLM2Rec consistently outperformed KAR and LLM-CF. This result validates the effectiveness of our distillation paradigm. Compared with KAR and LLM-CF, our distillation strategy effectively leverages the powerful recommendation capabilities of LLMs and directly transfers these merits to the student models. Compared with the Chain of Thought (CoT) utilized by LLM-CF, our distillation strategy directly utilizes the teacher's embeddings and recommendation results, which could be more easily digested by the student models.

\begin{table}[t]
\vspace{-0cm}
  \caption{Performance and efficiency comparison of BIGRec and DLLM2Rec on different datasets.}
  \vspace{-0.3cm}
  \label{tab:Efficiency}
  \setlength{\tabcolsep}{1mm}{
  \begin{tabular}{ccccc}
    \toprule
    Dataset&Model&HR@20&NDCG@20&Inference time\\
    \hline
    \multirow{3}{*}{Games}
    &BIGRec&0.0532&0.0341&2.3$\times10^4$s\\
    &DLLM2Rec&0.0751&0.0331&1.8s\\
    &\emph{Gain}&+37.41\%&-2.99\%&+1.3$\times10^6$\%\\
    \hline
    \multirow{3}{*}{MovieLens}
    &BIGRec&0.0541&0.0370&1.8$\times10^4$s\\
    &DLLM2Rec&0.1063&0.0437&1.7s\\
    &\emph{Gain}&+96.49\%&+18.18\%&+1.1$\times10^6$\%\\   
    \hline
    \multirow{3}{*}{Toys}
    &BIGRec&0.0420&0.0207&1.1$\times10^4$s\\
    &DLLM2Rec&0.0463&0.0225&1.6s\\
    &\emph{Gain}&+10.24\%&+8.70\%&+6.8$\times10^5$\%\\ 
  \bottomrule
\end{tabular}}
\end{table}

\textbf{Comparing with the teacher.} Table \ref{tab:Efficiency} shows the performance and efficiency comparison of the student model empowered by our DLLM2Rec with the teacher model BIGRec. To our surprise, the empowered lightweight student can even surpass the complex teacher model using LLMs. This result can be attributed to our design — we target letting the student digest the knowledge from the teacher while maintaining its own capacity to capture collaborative signals. Additionally, considering the inference efficiency, the empowered student still maintains low inference latency, while the BIGRec requires an unacceptably long inference time. This result validates that our DLLM2Rec can indeed address a crucial problem—maintaining excellent performance akin to LLM-based recommenders while ensuring low inference latency.

\begin{table}
  \caption{Overlapping ratio on Top-20 items.}
    \vspace{-0.3cm}
  \label{tab:overlapping}
  
  \begin{tabular}{ccccc}
    \toprule
     \textbf{Datasets} & \textbf{Before-distillation} & \textbf{Post-distillation}\\
    \hline    
    Games & 3.99\% & 10.88\% \\
    MovieLens & 4.06\% & 10.15\% \\
    Toys & 1.05\% & 14.56\%\\
  \bottomrule
  \end{tabular}
\end{table}

Besides, we conducted additional experiments to determine the ratio of overlapping items between the teacher and student models. As shown in Table \ref{tab:overlapping}, the post-distillation student model can effectively assimilate the teacher knowledge. Furthermore, it is important to note that the post-distillation student model may not entirely replicate the teacher's recommendations, given that the potential unreliability of teacher knowledge and the inherent teacher-student capacity gap.

\subsection{Ablation Study (RQ2)}
We conducted an ablation study on different datasets to study the contributions of each component of DLLM2Rec. For the importance-aware ranking distillation, we removed the entire component ($w/o \; all_r$), position-aware weights $w^p_{\mathbf si}$, confidence-aware weights $w^c_{\mathbf si}$, and consistency-aware weights $w^o_{\mathbf si}$, respectively. The results are presented in Table \ref{tab:Ablation}. For the collaborative embedding distillation, we removed the entire component ($w/o \; all_e$), the offset term, respectively, and tested the performance when replacing this module with existing embedding distillation strategies including Hint and HTD. The results are presented in Table \ref{tab:Ablation2}.

\begin{table}[t]
  \centering
  \caption{Ablation Study on ranking distillation.}
  \vspace{-0.3cm}
  \label{tab:Ablation}
  \scalebox{1.0}{
  \begin{tabular}{cccc}
    \toprule
    Dataset&Model&HR@20&NDCG@20 \\
    \hline
    \multirow{5}{*}{Games}
    &w/o $all_r$               &0.0661 	                &0.0301\\    
    &w/o $w^p_{\mathbf si}$            &0.0697 	                &0.0301\\
    &w/o $w^c_{\mathbf si}$            &0.0733 	                &0.0300\\
    &w/o $w^o_{\mathbf si}$            &0.0568 	                &0.0311\\
    &DLLM2Rec                        &\textbf{0.0751} 	    &\textbf{0.0331}\\
    \hline
    \multirow{5}{*}{MovieLens}
    &w/o $all_r$               &0.0917 	                &0.0364\\    
    &w/o $w^p_{\mathbf si}$            &0.1037	                &0.0429\\
    &w/o $w^c_{\mathbf si}$            &0.0986 	                &0.0398\\
    &w/o $w^o_{\mathbf si}$            &0.1047 	                &0.0430\\
    &DLLM2Rec                        &\textbf{0.1063} 	    &\textbf{0.0437}\\
    \hline
    \multirow{5}{*}{Toys}
    &w/o $all_r$               &0.0386 	                &0.0177\\    
    &w/o $w^p_{\mathbf si}$            &0.0406	                &0.0200\\
    &w/o $w^c_{\mathbf si}$            &0.0430 	                &0.0205\\
    &w/o $w^o_{\mathbf si}$            &0.0445 	                &0.0208\\
    &DLLM2Rec                        &\textbf{0.0463} 	    &\textbf{0.0225}\\
    \bottomrule
  \end{tabular}
  }
  \vspace{-0.3cm}
\end{table}

\begin{table}[t]
  \centering
  \caption{Ablation Study on embedding distillation.}
  \vspace{-0.3cm}
  \label{tab:Ablation2}
  \scalebox{1.0}{
  \begin{tabular}{cccc}
    \toprule
    Dataset&Model&HR@20&NDCG@20 \\
    \hline
    \multirow{5}{*}{Games}
    &w/o $all_e$                 &0.0649 	                &0.0323\\    
    &w/o \emph{offset}                      &0.0700 	                &0.0298\\
    &Hint                            &0.0563 	                &0.0244\\
    &HTD                             &0.0568 	                &0.0246\\
    &DLLM2Rec                        &\textbf{0.0751} 	    &\textbf{0.0331}\\
    \hline
    \multirow{5}{*}{MovieLens}
    &w/o $all_e$                 &0.0999 	                &0.0420\\    
    &w/o \emph{offset}                      &0.1061 	                &0.0425\\
    &Hint                            &0.0861 	                &0.0344\\
    &HTD                             &0.0874 	                &0.0341\\    
    &DLLM2Rec                        &\textbf{0.1063} 	    &\textbf{0.0437}\\
    \hline
    \multirow{5}{*}{Toys}
    &w/o $all_e$                 &0.0379 	                &0.0194\\    
    &w/o \emph{offset}                      &0.0405 	                &0.0195\\
    &Hint                            &0.0358 	                &0.0159\\
    &HTD                             &0.0349 	                &0.0157\\    
    &DLLM2Rec                        &\textbf{0.0463} 	    &\textbf{0.0225}\\
    \bottomrule
  \end{tabular}
  }
  \vspace{-0.3cm}
\end{table}

As can be seen, both distillation components are important — removing the importance-aware ranking distillation or the collaborative embedding distillation would result in performance drops. Delving deeper into the ranking distillation, we observe that developing the confidence-aware and consistency-aware weights are indeed helpful. For embedding distillation, we observe that the developed offset term is also important. More interestingly, by replacing the entire embedding distillation strategy with Hint and HTD, we observed quite poor performance. This could be attributed to the large semantic gap between the teacher and student models. Blindly aligning the embeddings may harm the model's semantic space. It could be even worse than directly inheriting the projected space from the teacher.

\subsection{Hyperparameter Sensitivity (RQ3)}
\begin{figure}[t]
\vspace{-0cm}
  \centering
  \includegraphics[width=0.48\textwidth]{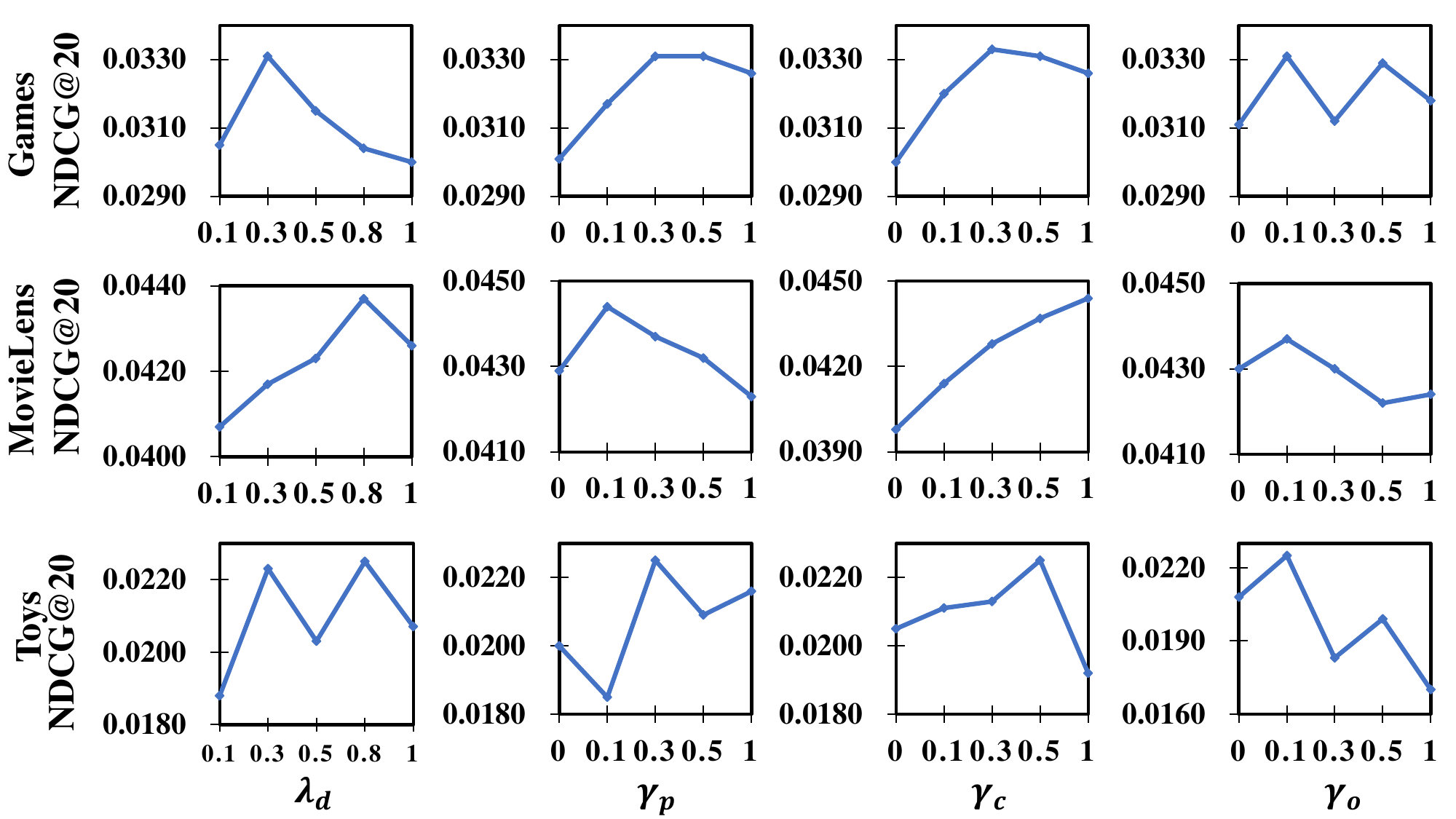}
  \vspace{-0.3cm}
  \setlength{\abovecaptionskip}{-0cm}
  \caption{Sensitivity analysis w.r.t. $\lambda_{d}$, $\gamma_{p}$, $\gamma_{c}$ and $\gamma_{o}$.}
  \label{fig:Sensitivity}
  \vspace{-2em}
\end{figure}

Figure \ref{fig:Sensitivity} illustrates performance of DLLM2Rec with different hyperparameters ($\lambda_d, \gamma_p, \gamma_c, \gamma_o$). While we observed some fluctuations, the general trend is that the model's performance would increase at the beginning and then drop as these parameters increase. This result validates the effectiveness of the corresponding components that each hyperparameter controls. But over-emphasizing one component would incur performance drops as it relatively declines the contribution from others. Finely tuning these hyperparameters for best balance could achieve optimal performance.

\section{RELATED WORK}
\subsection{LLMs for Recommendation System}
There are primarily two approaches to utilizing LLMs in RS: LLMs directly as recommenders \cite{bigrec, tallrec, li2023prompt, zhang2024prospect} and LLMs enhancing conventional recommenders \cite{yang2024common, xi2023towards, qin2024d2k}.

\textbf{LLMs as recommenders.} Initial efforts explored the zero-shot capabilities of LLMs in recommendation by structuring the recommendation tasks as language prompts \cite{gao2023chat, hou2024large, wang2023zero, liu2023first, wang2024recommend}. Subsequently, to adapt LLMs to recommendation tasks, instruction-tuning or fine-tuning has been widely adopted, showing promising results \cite{bigrec, tallrec, li2023prompt, zhang2023recommendation, li2024citationenhanced, wang2024efficient, hu2024exact, shi2024enhancing, li2023exploring, zhang2024tired, li2024paprec}. This research primarily focuses on how to enhance LLMs to better suit recommendation tasks. For instance, some studies aim to minimize the semantic gap between recommendations and natural language \cite{bigrec,zhu2023collaborative, hou2023learning, zheng2023adapting, wang2024enhanced, liao2023llara, xu2024enhancing, cao2024aligning, jiang2024item, tan2024llmrecsys}, such as the approach taken by BIGRec \cite{bigrec}, which uses a grounding strategy to map LLM descriptions to recommended items. Others focus on improving LLMs' ability to model long-sequence interactions \cite{geng2024breaking, zheng2024harnessing, lin2023rella}, identify noisy items \cite{wang2024llm4dsr} and some attempt to reduce training overhead \cite{lin2024data}. 
While effective, these methods often suffer from significant inference inefficiency, limiting their practical application. Although some studies have tried to mitigate inference latency through pre-storage techniques \cite{geng2024breaking} or knowledge distillation \cite{wang2024can}, the gain of \cite{geng2024breaking} is generally modest and \cite{wang2024can} is utilized to distill a huge LLMs (\eg GPT-115B) to a relatively smaller LLMs (\eg Llama-7B). Even small Llama-7B is hard to deploy in practical.

\textbf{LLMs enhancing conventional recommenders.} 
Existing methods mainly employ LLMs as supplementary knowledge bases \cite{yang2024common, xi2023towards, qin2024d2k} or as encoders for users/items \cite{ren2023representation, wei2024llmrec, wang2023alt} to improve conventional recommenders. For instance, KAR \cite{xi2023towards} utilizes LLMs as external knowledge bases to better profile users and items within the recommender system. RLMRec \cite{ren2023representation} encodes user and item profiles into semantic representations and aligns embeddings generated by conventional models with those from LLMs; Wei et al. \cite{wei2024llmrec} exploit LLMs to uncover new relationships within graphs. However, compared to direct LLM-based recommenders, these methods do not fully leverage the semantic reasoning capabilities of LLMs for making recommendations. CSRec \cite{yang2024common} incorporates the common sense extracted from LLM to alleviate the data sparsity issue.

To capitalize on the reasoning ability of teachers, some efforts have attempted to use chains of thought (CoT) data generated by LLMs to enhance conventional models \cite{sun2024large, wang2024can}. However, given the relatively simple architecture of conventional models, it is notably difficult for these systems to assimilate the complex knowledge from CoT data. Furthermore, CoT is typically used as a feature for a sequence, indicating that CoT for a new sequence should be produced during inference, which would be time-consuming. While some approaches attempt to retrieve similar CoT from other user-item pairs \cite{sun2024large}, such approximations may hurt accuracy. 

Differing from these methods, our distillation strategy capitalizes on the superiority of LLMs as recommenders, transferring their exceptional recommendation capabilities to conventional models. Our approach involves direct distillation on embeddings and ranking lists, which are easily assimilated by conventional models without incurring additional computational overhead during inference.

\subsection{Sequential Recommendation}
Sequential recommendation \cite{sun2019bert4rec, xie2022contrastive, chen2018sequential, chang2021sequential, li2020time} takes into account the sequence or order of interactions to predict what a user might prefer next. Existing Sequential RS use sequence generation models, such as RNNs \cite{hidasi2015session} or Transformers \cite{kang2018self, chen2024sigformer, dros}, to model user interaction sequences. For example, GRU4Rec \cite{hidasi2015session} employs the GRU to handle session-based data, while Caser \cite{tang2018personalized} uses CNN to model interaction data on multiple levels. SASRec \cite{kang2018self} introduces attention mechanism to automatically learn the weights of different interaction items, and DROS \cite{dros} leverage distributional robust optimization in sequential recommendation and achieves state-of-the-art performance. Some other methods focus on the intrinsic biases \cite{lin2024recommendation, chen2023bias, chen2021autodebias, zhao2022popularity} and distribution shifts \cite{wang2024distributionally} within RS. Most current LLM for RS methods also adopt the setting of sequential recommendation \cite{bigrec, tallrec}. The readers may refer to the excellent survey \cite{fang2020deep, wang2019sequential} for more details.

\subsection{Knowledge Distillation in RS}
Knowledge distillation (KD) is a promising model compression technique that transfers the knowledge from a large teacher model into the target compact student model \cite{close2open1}, and they have been widely applied in recommendation systems to reduce inference latency. RD \cite{rd} treated the top-N ranked items as positive for training a student model; CD \cite{cd} utilized soft labels to create positive and negative distillation instances; Soft labels  also considered by RRD \cite{rrd} to create the list-wise distillation loss function; DCD \cite{dcd} built distillation loss on both user-side and item-side; UnKD \cite{unkd} addresses popularity bias in distillation. The hidden knowledge among the middle layer of teachers are also considered in some methods. For example, Hint \cite{hinton2015distilling} and RRD \cite{rrd} extracted knowledge of teachers' embedding via Fitnet and expert network; HTD \cite{htd-21} distilled the topological knowledge with the relations in the teacher embedding space. Some researcher also study to distill knowledge from a huge LLMs (\eg ChatGPT) to a relatively smaller LLMs (\eg LLaMA-7B) in the recommendation scenarios.  
Besides model compression, KD is also used to integrate knowledge among different models \cite{kdingraph,kd-debias,kd-ensemble}. For example, some researchers \cite{kd-ensemble} consider to integrate knowledge from multiply pre-trained models into the student.   To the best of our knowledge, the study of distilling LLM-based recommenders into conventional recommenders remains untouched.

\section{CONCLUSION}
This work studies on distilling knowledge from LLM-based recommenders to conventional recommenders. The distillation encounters three challenges including potential unreliable teacher Knowledge, teacher-student capacity gap and semantic space divergence. To tackle this problem, we propose DLLM2Rec with leveraging \textit{importance-aware ranking distillation} and \textit{collaborative embedding distillation} for reliable and student-friendly distillation process. Extensive experiments demonstrate that DLLM2Rec can effectively enhance the performance of three typical lightweight conventional models, with an average improvement  of 47.97\%, enabling them to keep pace with sophisticated LLM-based models.

\begin{acks}
This work is supported by the National Natural Science Foundation of China (62372399) and the advanced computing resources provided by the Supercomputing Center of Hangzhou City University.
\end{acks}

\bibliographystyle{ACM-Reference-Format}
\bibliography{ref}


\end{document}